\documentclass[runningheads]{llncs}

\usepackage[numbers]{natbib}

\usepackage{float}

\usepackage[pdftex, pagebackref=true, pdfpagelabels=true,
hypertexnames=true, plainpages=false, naturalnames=false,
bookmarks=true, bookmarksnumbered=true,
pdfpagemode=UseOutlines,]{hyperref}
\usepackage{amsmath, multicol, multirow, booktabs, microtype, balance}
\usepackage{commath}

\hypersetup{colorlinks, citecolor=[rgb]{0,0.7,0}, filecolor=[rgb]{0.6,0.0,0.0}, linkcolor=[rgb]{0.0,0.0,0.5}, urlcolor=[rgb]{0.6,0.0,0}}

\usepackage{graphicx}\graphicspath{ {figures/} }

\begin{document}
\title{Window-Level is a Strong Denoising Surrogate}
%
\author{Ayaan Haque\inst{1,2}{} \and
Adam Wang\inst{2}{} \and
Abdullah-Al-Zubaer Imran\inst{2}{}}

\authorrunning{A. Haque et al.}
%
\institute{Saratoga High School, Saratoga, CA, USA \and
Stanford University, Stanford, CA, USA
}
\maketitle              
\begin{abstract}
CT image quality is heavily reliant on radiation dose, which causes a trade-off between radiation dose and image quality that affects the subsequent image-based diagnostic performance. However, high radiation can be harmful to both patients and operators. Several (deep learning-based) approaches have been attempted to denoise low dose images. However, those approaches require access to large training sets, specifically the full dose CT images for reference, which can often be difficult to obtain. Self-supervised learning is an emerging alternative for lowering the reference data requirement facilitating unsupervised learning. Currently available self-supervised CT denoising works are either dependent on foreign domain or pretexts are not very task-relevant. To tackle the aforementioned challenges, we propose a novel self-supervised learning approach, namely Self-Supervised Window-Leveling for Image DeNoising (SSWL-IDN), leveraging an innovative, task-relevant, simple, yet effective surrogate---prediction of the window-leveled equivalent. SSWL-IDN leverages residual learning and a hybrid loss combining perceptual loss and MSE, all incorporated in a VAE framework. Our extensive (in- and cross-domain) experimentation demonstrates the effectiveness of SSWL-IDN in aggressive denoising of CT (abdomen and chest) images acquired at 5\% dose level only.\footnote[1]{Code available at \url{https://github.com/zubaerimran/SSWL-IDN}}

\end{abstract}
\section{Introduction}

\begin{figure}[t]
    \centering
    \includegraphics[width=\linewidth, trim={0.5cm 0cm 0cm 0cm}, clip]{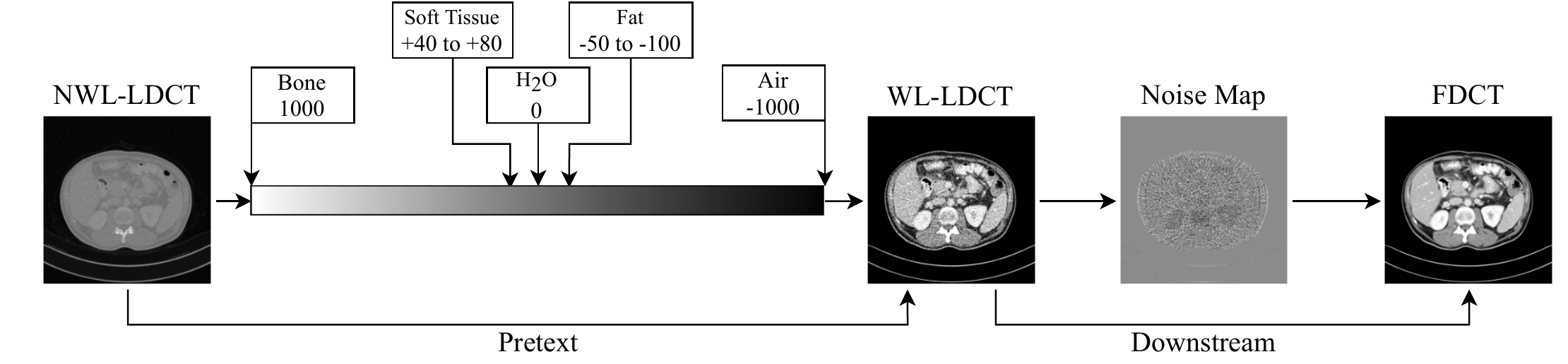}
    \caption{Window-Leveling is achieved by using the CT numbers and adjusting the contrast and brightness of the image. This image modification as a pretext, learns important representations of the data, facilitating improved downstream denoising of predicting FDCT from LDCT by removing noise.}
    \label{fig:task-figure}
\end{figure}

Computed Tomography (CT) imaging is one of the fundamental imaging modalities in medical practice. However, X-ray radiation is a clinical concern, as high radiation can be harmful to patients \citep{NEJMra072149}. Low dose CT (LDCT) images could be acquired to reduce radiation dose as an alternative to full dose CT (FDCT). However, lowering radiation dose introduces higher noise and various imaging artifacts, resulting in degraded diagnostic and other image-based performance. To address this tradeoff, deep learning-based \textit{denoising} methods have been investigated to improve and enhance CT imaging. Conventionally, denoising models map noisy LDCT (input) to cleaner FDCT (target). CT denoising is a popular field of research because of its clinical importance, as being able to denoise, and thus use, low dose CT provides improved patient safety and diagnostic performance.  Approaches include new architectures \citep{diwakar2018review, Chen_2017,Yang_2018, ataei2020cascaded} and training procedures \citep{chen2017low, yi2018sharpness}.

Acquiring reference images is challenging due to the harmful nature of radiation as well as the difficulty of performing two identical scans at different radiation doses. Thus, it is desirable to train denoising models with limited reference data. Self-Supervised Learning (SSL) has emerged as a promising alternative to fully-supervised learning in order to utilize large unlabeled training examples. In an SSL scheme, synthetic labels can be generated from the data itself, for both labeled and unlabeled data. Similar to transfer learning, SSL pre-trains a model on a surrogate task, but on the same dataset instead of one from a foreign domain, and then fine-tunes the pretrained model on a downstream, or main evaluation, task \citep{de1994learning}. Common surrogates available in literature include rotation prediction, colorization/restoration, and patch prediction. In general-purpose SSL denoising, popular works include \citep{xu2020noisy,laine2019high, Quan_2020_CVPR, xie2020noise2same}. In SSL CT denoising, less work exists \citep{krull2019noise2void, wu2020self, yuan2020half2half}.

Variational autoencoders (VAEs) \cite{kingma2013auto} are an extension of autoencoders (AEs), which use encoders and decoders to deconstruct inputs to low-dimensional representations and then reconstruct it. VAEs, however, are generative because they introduce the latent space where the latent code is injected with random noise. This is known as the \textit{reparameterization trick}. For denoising however, VAEs have not been extensively used \citep{im2017denoising, yue2019variational}. Additionally, residual learning \citep{he2016deep} has grown in interest as deep learning networks have become deeper, so Residual (ResNet) VAEs have been proposed for image generation \citep{kingma2016improving}. In medical image denoising, to our best knowledge, there is little literature using VAEs \citep{biswas2020dvae}, and the use of RVAEs are even more scarce. For loss functions, recent works support the use of Perceptual Loss because it optimizes high-level feature learning \citep{ataei2020cascaded, Yang_2018} as opposed to Mean-Squared Error (L2) which optimizes on a pixel-wise scale for precise noise removal. While few CT denoising methods have used hybrid losses \cite{ma2020low}, none have combined perceptual and pixel-wise losses.

In this paper, we use SSL to improve performance of deep denoising models with limited reference FDCT. We propose a novel denoising surrogate to predict \textit{window-leveled} CT images from non-window-leveled images as a pretext. Window-leveling in CT is the process of modifying the grayscale of an image, using the CT numbers, to highlight, brighten, and contrast important structures. 
Unlike many other existing self-supervised learning methods, our proposed self-supervised window-leveling (SSWL) is a task-relevant surrogate, as it is directly related to the downstream task, prioritizing similar feature learning.
Furthermore, we limit all our experiments to 5\% dose level potentially towards an aggressive dose reduction mechanism and demonstrate effectiveness even at such low dose settings.

Our primary contributions can be summarized as follows:

\begin{itemize}
\item 
A novel and task-relevant self-supervised window-level prediction surrogate which is related to the downstream task.

\item 
An innovative residual-based VAE architecture coupled with a hybrid loss function to simultaneously penalize the model pixel-wise and perceptually.

\item 
Extensive experimentation with varied quantities of labeled data on different proposed components on in- and cross-domain data demonstrating improved and effective denoising even from extremely low dose (5\%) CT images.
\end{itemize}

\section{Methods}
\label{sec:method}

\begin{figure}[t]
    \centering
    \includegraphics[width=0.9\linewidth]{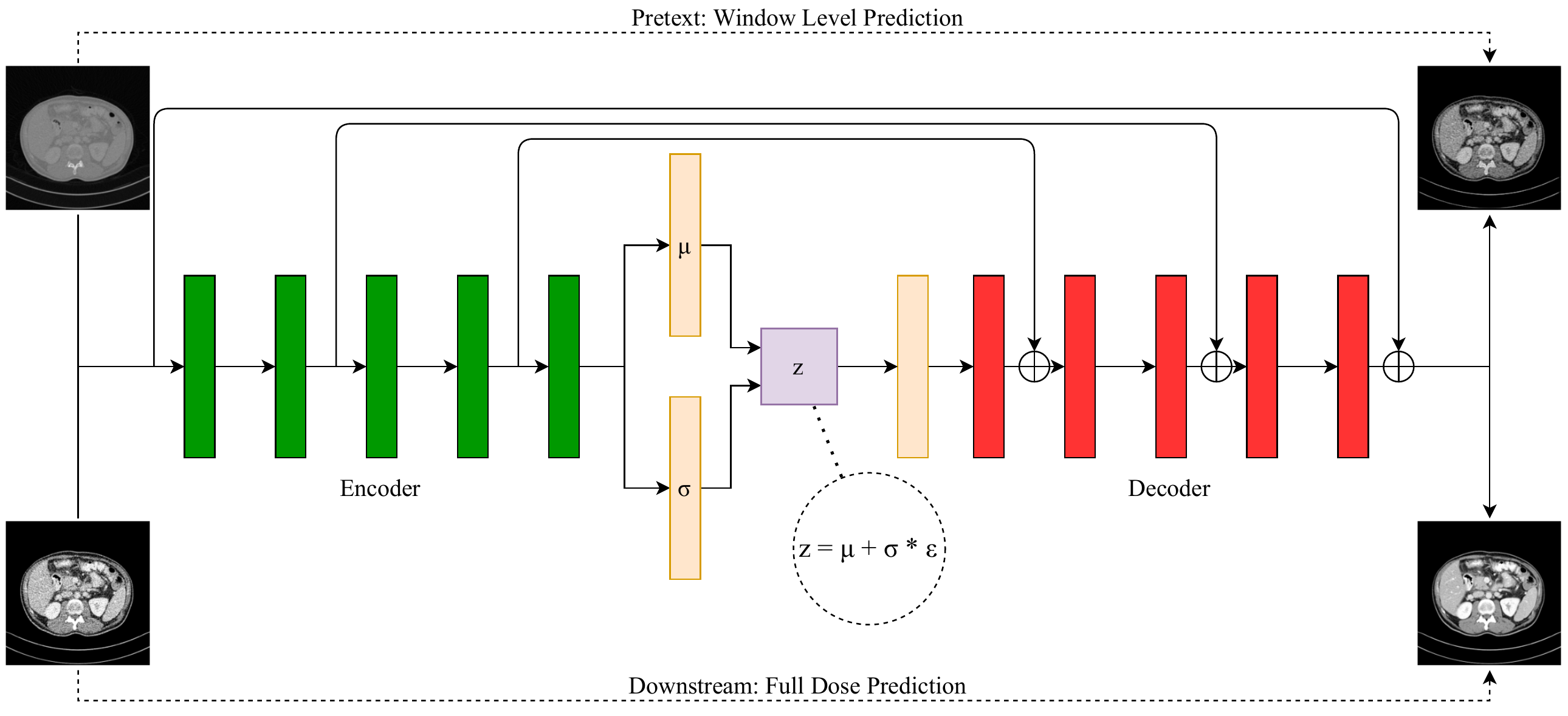}
    \caption{Schematic of the proposed SSWL-IDN model. For the SSL surrogate task, the model predicts window-leveled images from non-window-leveled images. For the downstream task, the model denoises the input LDCT to match the FDCT. The model architecture uses residuals between the encoder and decoder, and the VAE bottleneck is incorporated with the reparameterization trick.}
    \label{fig:model}
\end{figure}

\subsection{Denoising}

To formulate the problem, we assume unknown data distribution $p(X,Y)$ over LDCT $X$ and FDCT $Y$. We also assume access to labeled training set $\mathcal{D}_l$ sampled i.i.d.~from $p(X,Y)$ and unlabeled training set $\mathcal{D}_u$ sampled i.i.d.~from $p(X)$ after marginalizing out $Y$. In CT denoising, the input images are LDCT and the reference images are FDCT. This relationship can be represented by the equation 

\begin{equation}
    X = Y + n%
\end{equation}
where $n$ is the resultant noise due to lowering dose. The deep denoising model is trained to remove $n$ by encoding the input LDCT and recovering the FDCT.

A similar relationship can be found in the CT window-leveling task---non-window-leveled (NWL) inputs for LDCT and window-leveled (WL) inputs for FDCT. Therefore, 

\begin{equation}
    Z = aX + b,
\end{equation}
where $X$ is NWL, $Z$ is WL, and $a$ and $b$ are window-leveling parameters. Fig.~\ref{fig:task-figure} illustrates the process of window-leveling and its relation to denoising. Inspired from the aforementioned relatedness of the two tasks, we can leverage the first task to help learn or improve the second. The window-leveling labels freely available in the CT metadata enables us to train a denoising deep learning model as if it is fully-supervised. Specifically, formulating it as a pretext to the downstream denoising task is more appropriate when obtaining full dose reference images is extremely difficult. Especially since the task is domain-specific, it allows for more important and relevant feature learning than foreign or arbitrary surrogates. Therefore, our proposed self-supervised learning method comprises of two steps: fully-supervised pre-training on the window-leveling task followed by fine-tuning on the small labeled denoising task. For pre-training, we prepare both a NWL and WL version of each LDCT scan for both labeled and unlabeled data. Loss is optimized for predicting the WL LDCT from input NWL LDCT. Our surrogate is end-to-end as opposed to many other methods which do not use related tasks, as no architectural or loss changes are required between tasks.

\subsection{Model Architecture and Training}

For the model architecture, illustrated in Figure \ref{fig:model} along with the SSL algorithm, we propose a Residual Variational Autoencoder (RVAE), which is a combination of \citep{Chen_2017} and \citep{kingma2013auto}. While Residual-based VAEs have been proposed \citep{kingma2016improving}, they rather use residuals in the encoder and decoder separately (ResNet as encoder and Transposed ResNet as decoder) instead of using residual connections \textit{between} the encoder and decoder like \citep{Chen_2017}. Additionally, to our knowledge, none have been used for medical image denoising in particular.

We use the base architecture of \citep{Chen_2017} and add a bottleneck component with Global Average Pooling and Linear Layers. This downsamples the input to a latent representation, where we use the reparameterization trick to adjust the latent code. This improves FD predictions, as adding randomized noise $\epsilon$, which is tunable through learnable parameters $\mu$ (mean) and $\sigma$ (standard deviation), in the bottleneck can decrease overfitting, improve generalization, and act as a regularizer. To reparameterize, we use the calculation $z = \mu + \sigma * \epsilon$. Additionally, a generative model, as opposed to standard AEs, can allow for better FD predictions, as the noise in LDCT may hide important details and features which can be more easily recovered through generative models, as denoising tends to overmsooth and remove subtle information. As opposed to traditional VAEs, we do not use downsampling convolutional layers but rather use constant convolutional filters of 96. We use residuals from feature maps in the encoder and add them after corresponding layers in the decoder phase. We use standard convolutional layers in the encoder and transpose convolutional layers in the decoder.

For our loss objective, we propose a hybrid loss combining mean-squared-error (MSE) and perceptual loss. Perceptual loss encourages high-level feature learning/matching (visual features), while MSE allows for precise, pixel-wise noise removal. To leverage the benefits of both, we combine them into a single function. We use \citep{zhang2018perceptual}'s definition of perceptual loss, where the model prediction and target are passed to a Image-Net pre-trained VGG-19 network and the predicted features from hidden convolutional layers are used to determine the perceptual distance between the images. The perceptual loss can be defined as 

\begin{equation}
    \mathcal{L}_{perceptual} = \frac{1}{M}\sum_{i=1}^{M}\norm{\mathcal{F}(\hat{y}_{i}) - \mathcal{F}(y_{i})}^{2},
\end{equation}

where $M$ is the mini-batch size, $\hat{y}$ are the model predictions, $y$ are the labels, and $\mathcal{F}$ is a feature extractor.
f
Our final loss function can be represented by 
\begin{equation}
    L(y, \hat{y}, \mu, \sigma) = L_{MSE}(\hat{y}, y) + \beta\mathcal{L}_{perceptual}(\mathcal{F}(\hat{y}), \mathcal{F}(y)) + \alpha\mathcal{L}_{KL}(\mu, \sigma),
\end{equation}

where $L_{MSE}$ is standard MSE loss, the $\mathcal{L}_{perceptual}$ is perceptual loss, and $\beta$ is $\mathcal{L}_{perceptual}$ weight. For the VAE, $\mathcal{L}_{KL}$ represents the KL divergence loss, and $\alpha$ is the $\mathcal{L}_{KL}$ weight. $\mu$ is the mean term, and $\sigma$ is the standard deviation term, both from the latent space. The KL divergence attempts to reduce divergence of the two parameters from the parameters of the target distribution.

\section{Empirical Evaluation}

\subsection{Data}

We primarily collect abdomen scans from the publicly available Mayo CT data \citep{chen2015development, yu2012development}. 
The dataset includes CT scans which are originally acquired at routine dose level (full dose), so simulated quarter dose images are reconstructed through inserting Poisson noise into each projection dataset. The images are window-leveled with width 300 and center 40. For thorough denoising evaluation, we generate the CT scans at 5\% dose level using the full dose and quarter dose data (scaling the zero-mean independent noise from 25\% to 5\% dose level) \citep{imran2021ssiqa}. While from a clinical perspective it is more ideal to have a well-denoised quarter dose as opposed to a lower quality denoised 5\% dose, from a computational perspective, showing the ability to remove high volumes of noise can more appropriately evaluate the model's full potential to accurately remove noise. We use 15 full dose abdomen CT and the corresponding quarter (25\%) dose CT scans, and split them into 10 scans (1,533 slices, 15\% used for validation set) for training and 5 (633 slices) for testing. Furthermore, 5 chest scans (1,061 slices) are selected from the same library for cross-domain evaluation. For chest scans, the 5\% dose level is simulated from the routine and 10\% dose level scans available in the Mayo data library.

\subsection{Implementation Details}
\textbf{Baselines:} For architectural baselines, we used a VAE \citep{kingma2013auto}, U-Net \citep{ronneberger2015u}, DnCNN \citep{zhang2017beyond}, and RED-CNN \citep{Chen_2017}. For baseline loss functions, we used MSE and perceptual loss. For SSL surrogate task baselines, we used no SSL, simple reconstruction (Rec), and a recent work namely Noisy-as-Clean (NAC) \citep{xu2020noisy}. \textbf{Training:} Models  were trained at varying levels of supervision in terms of the number of labeled data (250, 500, 1000, full). Each experiment was performed 5 times, and the means were reported. All inputs were normalized and resized to 256 $\times$ 256 $\times$ 1. Inference takes 0.01 seconds, and fully-supervised training takes approximately 3.4 min/epoch.
\textbf{Hyperparameters:} The code was written in the most recent versions of Python and PyTorch and were trained on an NVIDIA K80 GPU with 12GB RAM. We used the Adam optimizer with learning rates of 1e-5 and momentum of 0.1 per 8 epochs. Each model was trained with a minibatch size 10. For loss, $\beta$ = 0.6 (through tuning experiments), and $\alpha$ = 1.0 (as per VAEs). Based on all the tests, the choice of hyperparameters do not affect performance much. \textbf{Evaluation:} For evaluation, we used Peak Signal-to-Noise Ratio (PSNR), Structural Similarity Index Measure (SSIM), Mean-Squared Error (MSE), and Normalized RMSE (NRMSE). To evaluate statistical significance, we performed two-sample t-tests.

\subsection{Results \& Discussion}

\begin{table}
\centering
\setlength{\tabcolsep}{4pt}
\caption{Our RVAE has significantly stronger performance than baseline architectures (trained on MSE) on in- and cross-domain evaluation. The best fully-supervised scores are bolded, and best semi-supervised scores are underlined.}
\label{table:archi-scores}
\resizebox{0.9\linewidth}{!}{
\begin{tabular}{@{}lc c cccc ccccc@{}}
            \toprule
          \multirow{3}{*}{Model}
           &
           \multirow{3}{*}{$|\mathcal{D}_l|$}
           &
           \multicolumn{4}{c}{In-Domain}
           &&
           \multicolumn{4}{c}{Cross-Domain}
          \\
          \cmidrule{3-6}\cmidrule{8-11}
          && PSNR & SSIM & MSE & NRMSE && PSNR & SSIM & MSE & NRMSE \\
          \midrule
           \multirow{1}{*}{LDCT}
           &
           ---
           &
           20.179 & 0.7554 & 0.0107 & 0.3082
           &&
           16.876 & 0.6725 & 0.0207 & 0.4832
           \\
           \midrule
           \multirow{4}{*}{U-Net}
           &
           250
           &
           17.325 & 0.5685 & 0.0201 & 0.4204
           &&
           16.434 & 0.4781 & 0.0234 & 0.5186
           \\
           &
           500
           &
           18.445 & 0.6606 & 0.0144 & 0.3629
           &&
           17.385 & 0.6094 & 0.0187 & 0.4647
           \\
           &
           1000
           &
           20.930 & 0.7697 & 0.0083 & 0.2746
           &&
           18.458 & 0.6883 & 0.0147 & 0.4110
           \\
           &
           Full
           &
           21.900 & 0.7918 & 0.0067 & 0.2470
           &&
           18.843 & 0.7095 & 0.0135 & 0.3933
           \\
           \midrule
           \multirow{4}{*}{VAE}
           &
           250
           &
           17.278 & 0.4656 & 0.0192 & 0.4173
           &&
           16.318 & 0.3741 & 0.0238 & 0.5245
           \\
           &
           500
           &
           19.241 & 0.5849 & 0.0121 & 0.3324
           &&
           17.329 & 0.4859 & 0.0188 & 0.4671
           \\
           &
           1000
           &
           20.771 & 0.6873 & 0.0086 & 0.2802
           &&
           18.094 & 0.5634 & 0.0148 & 0.4266
           \\
           &
           Full
           &
           21.547 & 0.7510 & 0.0073 & 0.2573
           &&
           18.364 & 0.5971 & 0.0145 & 0.4091
           \\
           \midrule
           \multirow{4}{*}{DnCNN}
           &
           250
           &
           19.814 & 0.7031 & 0.0112 & 0.3171
           &&
           16.861 & 0.6132 & 0.0207 & 0.4777
           \\
           &
           500
           &
           20.430 & 0.7372 & 0.0101 & 0.2981
           &&
           17.192 & 0.6554 & 0.0202 & 0.4740
           \\
           &
           1000
           &
           21.693 & 0.7460 & 0.0071 & 0.2533
           &&
           18.702 & 0.6343 & 0.0157 & 0.4217
           \\
           &
           Full
           &
           21.186 & 0.7907 & 0.0088 & 0.2749
           &&
           18.925 & 0.6584 & 0.0136 & 0.3662
           \\
           \midrule
           \multirow{4}{*}{RED-CNN}
           &
           250
           &
           23.414 & 0.8433 & 0.0054 & 0.2155
           &&
           18.625 & 0.6971 & 0.0139 & 0.4319
           \\
           &
           500
           &
           23.612 & 0.8522 & 0.0051 & 0.2078
           &&
           18.919 & 0.7264 & 0.0132 & 0.4019
           \\
           &
           1000
           &
           24.097 & 0.8538 & 0.0052 & 0.1990
           &&
           19.005 & 0.7327 & 0.0129 & 0.3819
           \\
           &
           Full
           &
           24.115 & 0.8616 & 0.0047 & 0.1979
           &&
           19.005 & 0.7467 & 0.0124 & 0.3745
           \\
           \midrule
           \multirow{4}{*}{RVAE}
           &
           250
           &
           24.829 & 0.8483 & 0.0053 & 0.1889
           &&
           19.033 & 0.7438 & 0.0131 & 0.3861
           \\
           &
           500
           &
           26.139 & 0.8538 & 0.0050 & 0.1889
           &&
           19.193 & 0.7467 & 0.0127 & 0.3789
           \\
           &
           1000
           &
           \underline{26.286} & \underline{0.8626} & \underline{0.0046} & \underline{0.1878}
           &&
           \underline{19.259} & \underline{0.7497} & \underline{0.0125} & \underline{0.3763}
           \\
           &
           Full
           &
           \textbf{26.574} & \textbf{0.8646} & \textbf{0.0045} & \textbf{0.1817}
           &&
           \textbf{19.288} & \textbf{0.7490} & \textbf{0.0122} & \textbf{0.3690}
           \\
           \bottomrule
    \end{tabular}}
\end{table}

Table \ref{table:archi-scores} shows our proposed RVAE, without SSL or hybrid loss, is able to outperform all baselines, including two state-of-the-art denoising models in DnCNN \citep{zhang2017beyond} and RED-CNN \citep{Chen_2017}, and even with minimum data, we are able to match the performance of fully-supervised versions of those two architectures. Our RVAE outperforms RED-CNN with statistical significance ($p$-value of 0.032 for PSNR comparison), and similarly significantly outperforms the other models ($p$\textless  0.05). Similarly for the cross-domain chest dataset evaluation, similar improvements are shown from RVAE compared to the baselines and the state-of-the-art, proving the generalizatibility of our model.

Table \ref{table:ssl-scores} displays the performance of DnCNN, RED-CNN, and our RVAE with various SSL tasks (trained on MSE). SSL Reconstruction uses the LDCT as the input and reference. For the Noisy-As-Clean surrogate task \cite{xu2020noisy}, we follow the authors' implementation, where their SSL task uses LDCT that is injected with additional Gaussian noise as the input, while the original LDCT is used as the reference. As shown in the tables, SSWL has significantly improved performance compared to no pre-training as well as basic reconstruction ($p$-value of 0.021 and 0.039 respectively).
When compared to NAC, the state-of-the-art, we still have the best performance, confirming the importance and ability of our proposed SSL task. Similar to in-domain evaluations, the cross-domain metrics further confirm the improved performance and generalization of SSWL over other methods.



\begin{figure}
    \centering
    \resizebox{0.95\linewidth}{!}{%
      \begin{tabular}{cccccc}
        \\
        
        \includegraphics[width=0.5\linewidth, trim={3.97cm 1.265cm 3.58cm 1.22cm},clip]{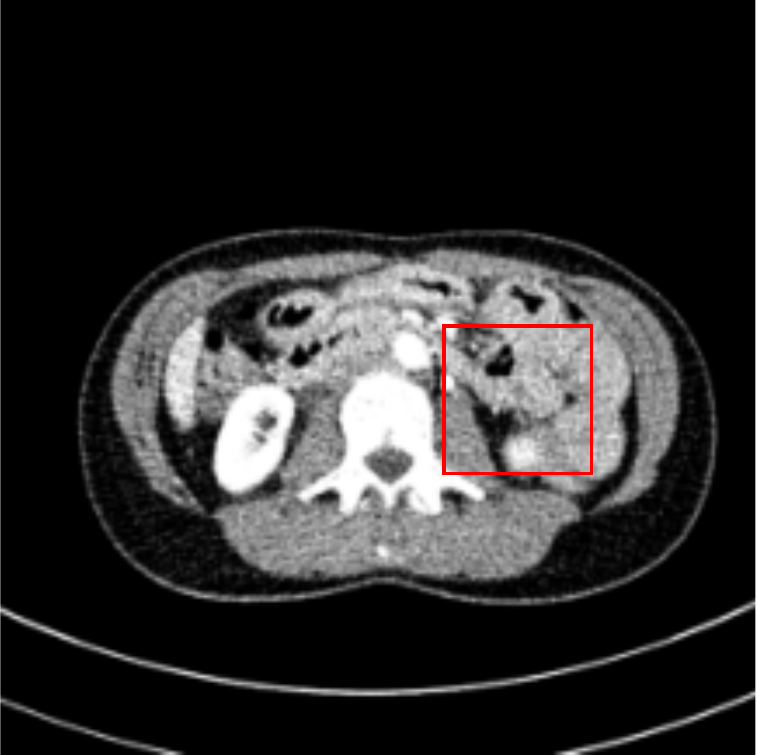}
        &
        \includegraphics[width=0.5\linewidth, trim={3.97cm 1.265cm 3.58cm 1.22cm},clip]{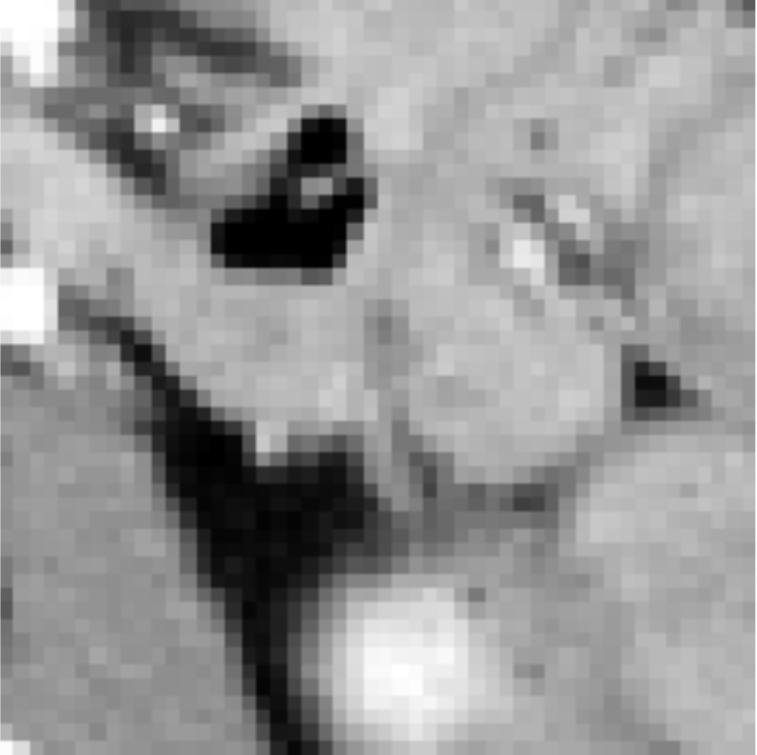}
        &
        \includegraphics[width=0.5\linewidth, trim={3.97cm 1.265cm 3.58cm 1.22cm},clip]{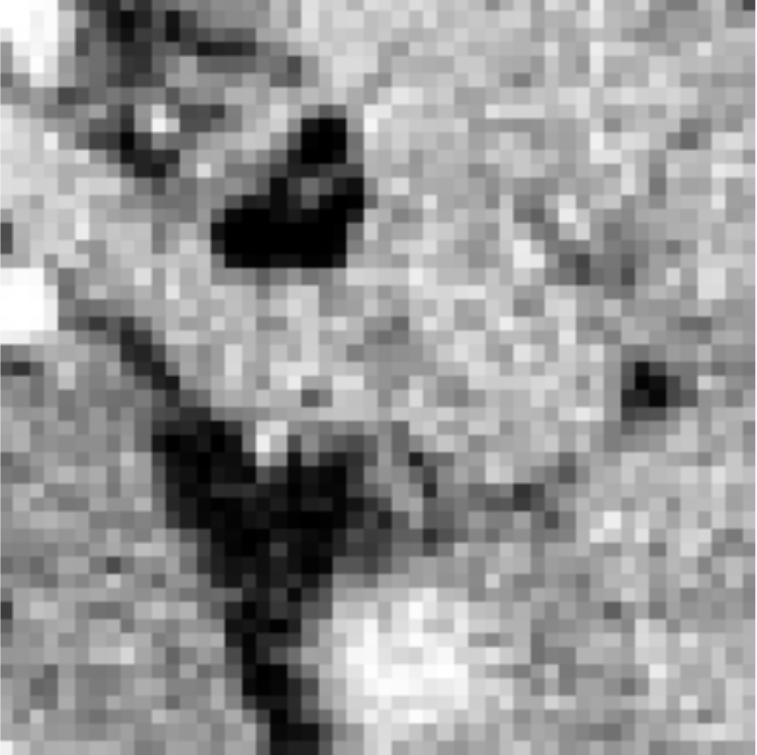}
        &
        \includegraphics[width=0.5\linewidth, trim={3.97cm 1.265cm 3.58cm 1.22cm},clip]{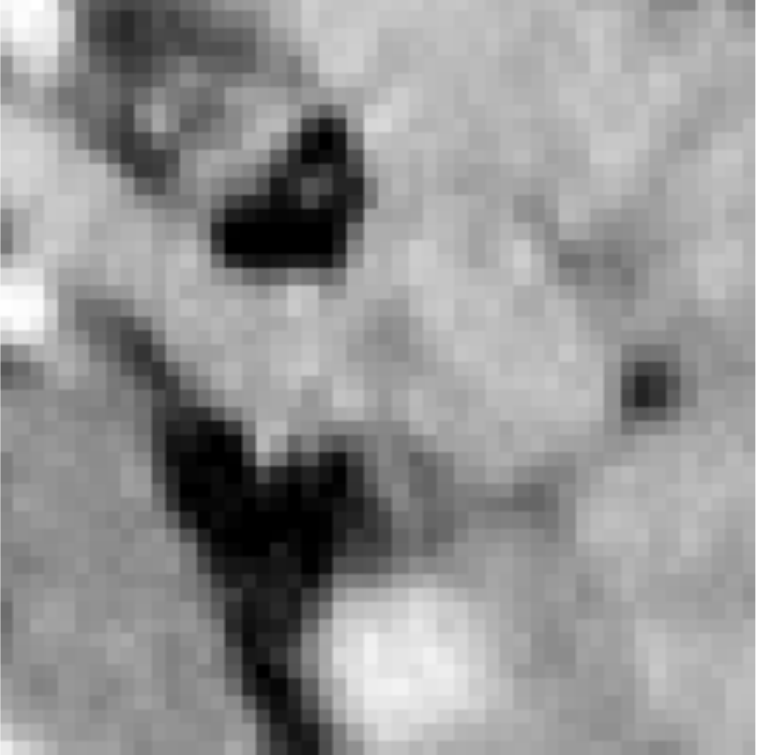}
        &
        \includegraphics[width=0.5\linewidth, trim={3.97cm 1.265cm 3.58cm 1.22cm},clip]{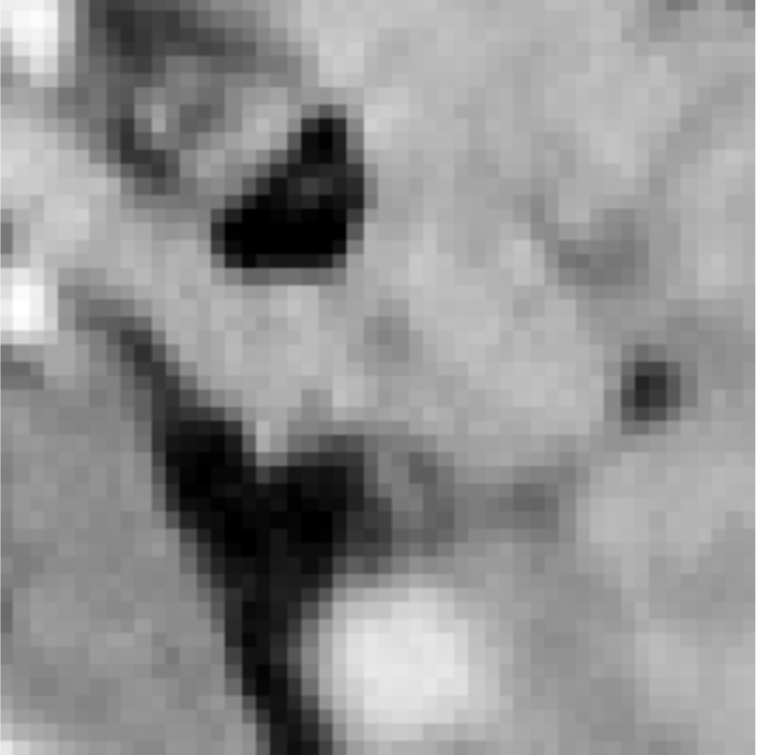}
        &
        \includegraphics[width=0.5\linewidth, trim={3.97cm 1.265cm 3.58cm 1.22cm},clip]{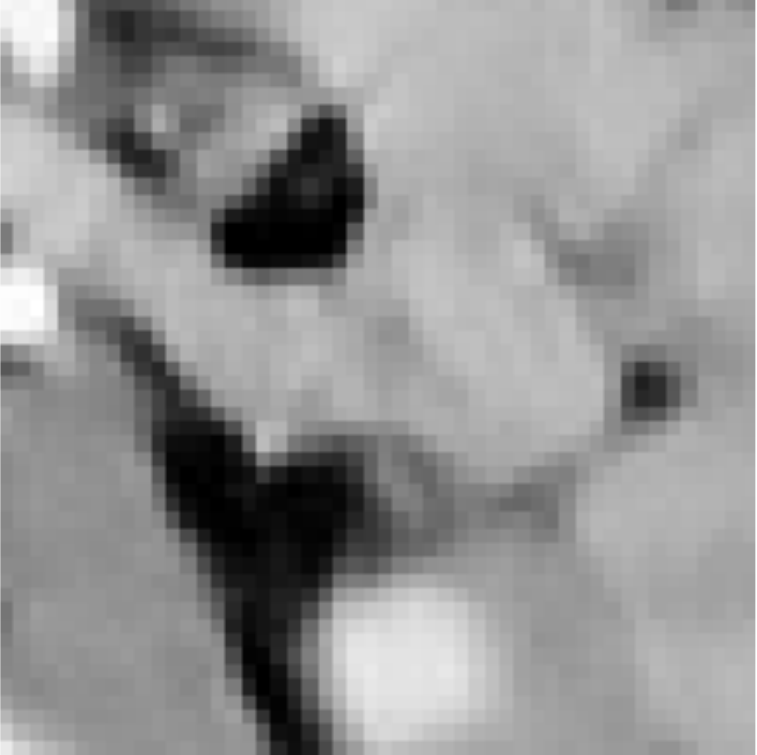}
        \\
        \smallskip
        {\Huge Image} & {\Huge GT ROI} & {\Huge LDCT (0.7675)} & {\Huge REC (0.8756)} & {\Huge NAC (0.8802)} & {\Huge SSWL (0.8884)}
        \end{tabular} }
    \caption{ROI predictions with RVAE ($|\mathcal{D}_l|$=500) and the 3 SSL tasks show higher visual noise removal and SSIM from SSWL.}
    \label{fig:comparison-ssl-rois}
\end{figure}

\begin{figure}
    \centering
    \resizebox{0.8\linewidth}{!}{%
      \begin{tabular}{ccc}
        \\
        
        \includegraphics[width=0.5\linewidth, trim={4.25cm 2.1cm 3.75cm 2.1cm},clip]{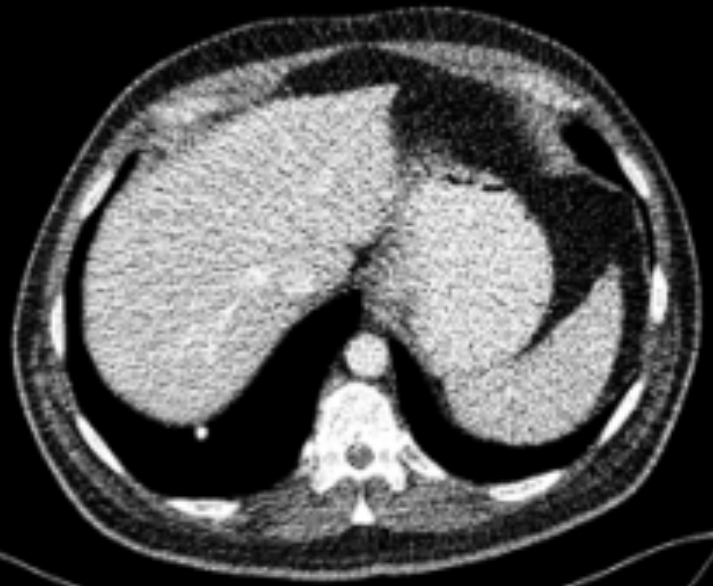}
        &
        \includegraphics[width=0.5\linewidth, trim={4.25cm 2.1cm 3.75cm 2.1cm},clip]{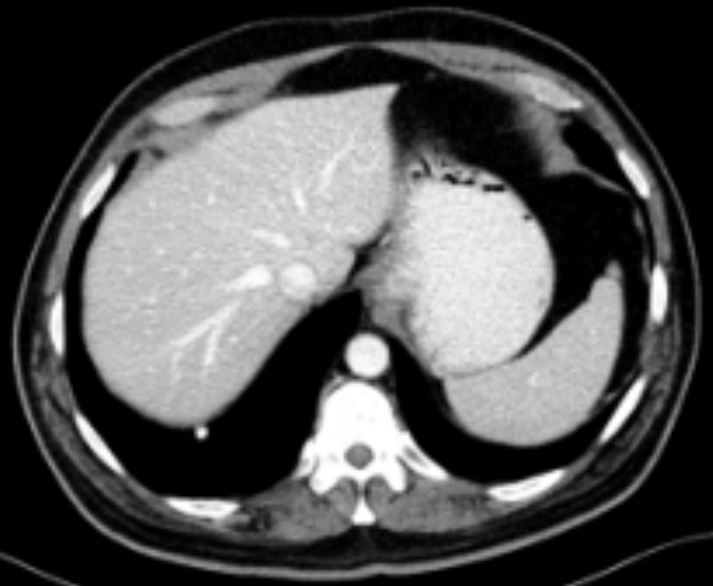} 
        &
        \includegraphics[width=0.5\linewidth, trim={4.25cm 2.1cm 3.75cm 2.1cm},clip]{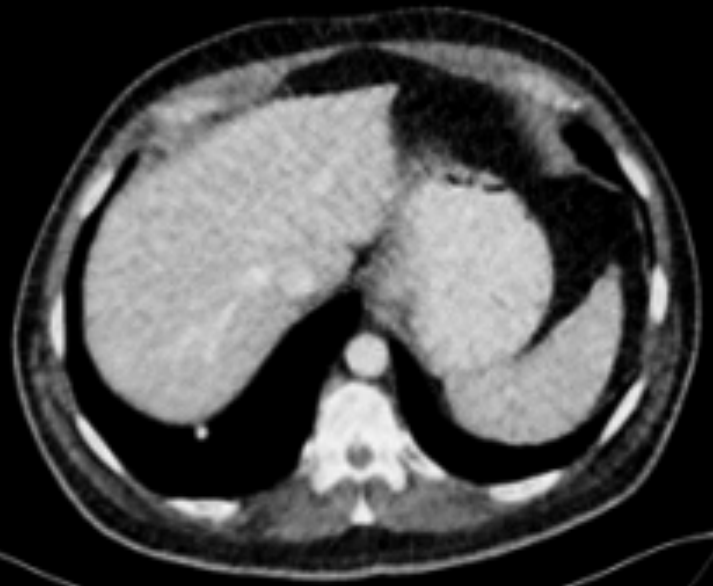}
        \smallskip \\
        {\huge LDCT (0.7586)} & {\huge FDCT} & {\huge RED-CNN (0.8790)}
         \bigskip\\
        \includegraphics[width=0.5\linewidth, trim={4.25cm 2.1cm 3.75cm 2.1cm},clip]{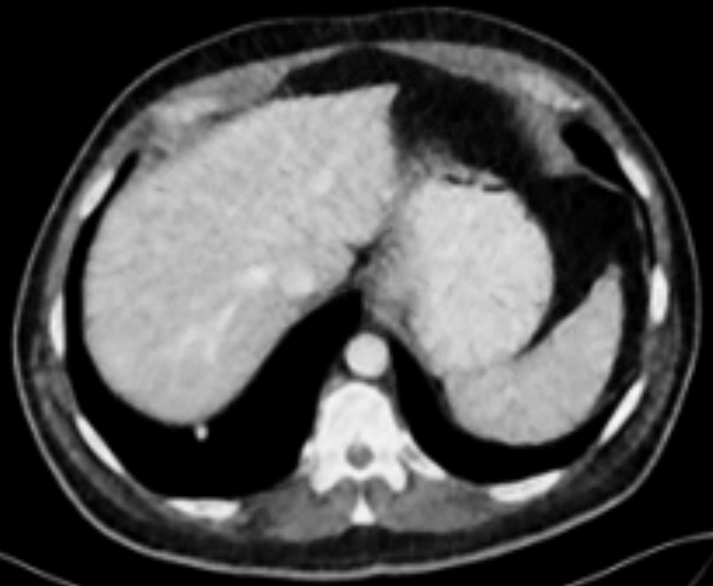}
        &
        \includegraphics[width=0.5\linewidth, trim={4.25cm 2.1cm 3.75cm 2.1cm},clip]{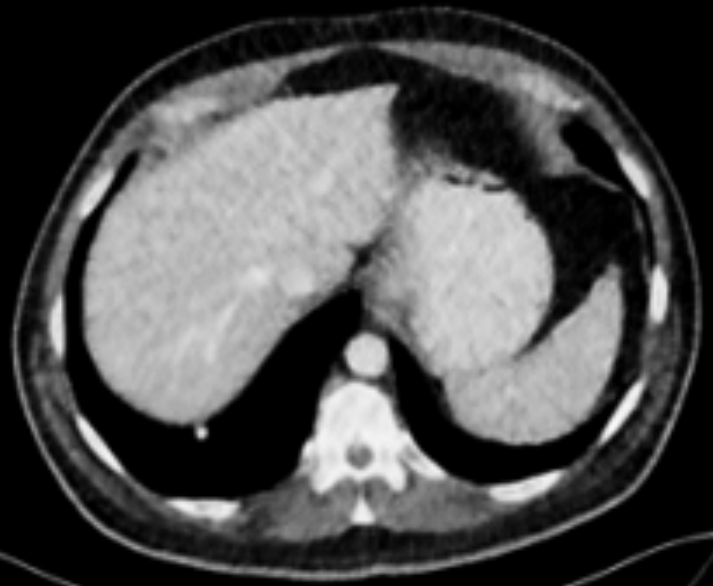}
        &
        \includegraphics[width=0.5\linewidth, trim={4.25cm 2.1cm 3.75cm 2.1cm},clip]{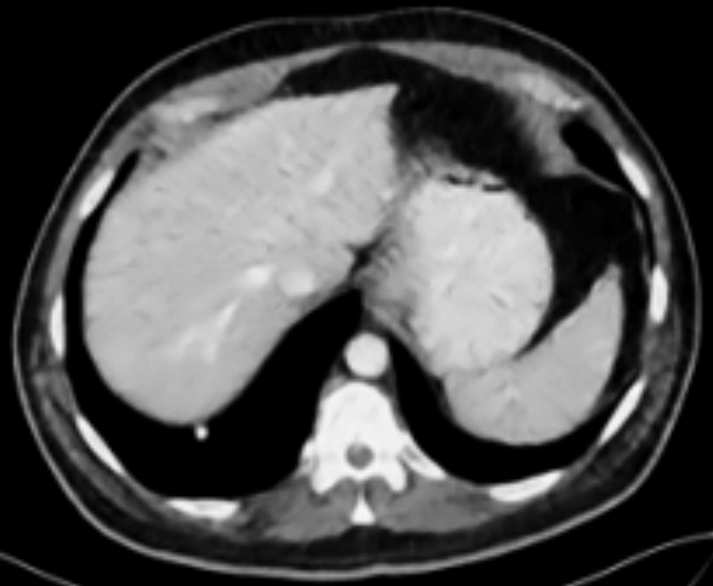}
        \smallskip\\
         {\huge RED-CNN+SSWL (0.8901)} & {\huge RVAE (0.8851)} & {\huge RVAE+SSWL (0.9023)}
        \end{tabular} }
        \caption{FDCT predictions from RED-CNN and RVAE ($|\mathcal{D}_l|$=500) with and without SSWL show strong visual improvements and improved SSIM from our proposed models.}
        \label{fig:comparison-preds}
\end{figure}



\begin{table}
\centering
\setlength{\tabcolsep}{4pt}
\caption{Metrics of SSWL are superior to Rec and NAC RED-CNN and RVAE architectures for in- and cross-domain evaluations. Scores are annotated like Table \ref{table:archi-scores}.}
\label{table:ssl-scores}
\resizebox{0.95\linewidth}{!}{
\begin{tabular}{@{}lc c cccc ccccc@{}}
            \toprule
          \multirow{3}{*}{Model}
           &
           \multirow{3}{*}{$|\mathcal{D}_l|$}
           &
           \multicolumn{4}{c}{In-Domain}
           &&
           \multicolumn{4}{c}{Cross-Domain}
          \\
          \cmidrule{3-6}\cmidrule{8-11}
          && PSNR & SSIM & MSE & NRMSE && PSNR & SSIM & MSE & NRMSE \\
           \midrule
           \multirow{4}{*}{RED-CNN + REC}
           &
           250
           &
           23.414 & 0.8433 & 0.0054 & 0.2155
           &&
           19.351 & 0.7551 & 0.0122 & 0.3726
           \\
           &
           500
           &
           23.612 & 0.8522 & 0.0051 & 0.2078
           &&
           19.430 & 0.7572 & 0.0120 & 0.3691
           \\
           &
           1000
           &
           24.097 & 0.8538 & 0.0052 & 0.1990
           &&
           19.499 & 0.7523 & 0.0117 & 0.3644
           \\
           &
           Full
           &
           24.115 & 0.8616 & 0.0047 & 0.1979
           &&
           19.546 & 0.7555 & 0.0120 & 0.3678
           \\
           \midrule
           \multirow{4}{*}{RVAE + REC}
           &
           250
           &
           23.795 & 0.8522 & 0.0051 & 0.2120
           &&
           19.417 & 0.7583 & 0.0120 & 0.3694
           \\
           &
           500
           &
           24.088 & 0.8598 & 0.0050 & 0.2067
           &&
           19.544 & 0.7602 & 0.0117 & 0.3640
           \\
           &
           1000
           &
           24.071 & 0.8650 & 0.0048 & 0.2007
           &&
           19.570 & 0.7598 & 0.0116 & 0.3632
           \\
           &
           Full
           &
           24.301 & 0.8669 & 0.0047 & 0.1958
           &&
           19.609 & 0.7613 & 0.0115 & 0.3614
           \\
           \midrule
           \multirow{4}{*}{RED-CNN + NAC}
           &
           250
           &
           23.989 & 0.8375 & 0.0050 & 0.2050
           &&
           19.122 & 0.7527 & 0.0124 & 0.3750
           \\
           &
           500
           &
           24.033 & 0.8541 & 0.0048 & 0.2015
           &&
           19.401 & 0.7565 & 0.0127 & 0.3703
           \\
           &
           1000
           &
           24.219 & 0.8549 & 0.0045 & 0.1961
           &&
           19.494 & 0.7577 & 0.0118 & 0.3660
           \\
           &
           Full
           &
           24.422 & 0.8611 & 0.0045 & 0.1958
           &&
           19.579 & 0.7588 & 0.0116 & 0.3625
           \\
           \midrule
           \multirow{4}{*}{RVAE + NAC}
           &
           250
           &
           23.890 & 0.8612 & 0.0052 & 0.2029
           &&
           19.313 & 0.7545 & 0.0123 & 0.3739
           \\
           &
           500
           &
           24.168 & 0.8579 & 0.0047 & 0.1996
           &&
           19.338 & 0.7532 & 0.0123 & 0.3726
           \\
           &
           1000
           &
           24.019 & 0.8647 & 0.0047 & 0.1980
           &&
           19.439 & 0.7580 & 0.0122 & 0.3686
           \\
           &
           Full
           &
           24.186 & 0.8649 & 0.0046 & 0.1956
           &&
           19.520 & 0.7596 & 0.0118 & 0.3650
           \\
           \midrule
           \multirow{4}{*}{RED-CNN + SSWL}
           &
           250
           &
           26.119 & 0.8509 & 0.0050 & 0.1890
           &&
           19.550 & 0.7611 & 0.0117 & 0.3635
           \\
           &
           500
           &
           26.300 & 0.8542 & 0.0050 & 0.1865
           &&
           19.460 & 0.7614 & 0.0120 & 0.3679
           \\
           &
           1000
           &
           26.710 & 0.8627 & 0.0046 & 0.1786
           &&
           19.520 & 0.7617 & 0.0120 & 0.3652
           \\
           &
           Full
           &
           26.747 & 0.8626 & 0.0045 & 0.1764
           &&
           19.547 & 0.7619 & 0.0117 & 0.3642
           \\
           \midrule
           \multirow{4}{*}{RVAE + SSWL}
           &
           250
           &
           26.150 & 0.8612 & 0.0051 & 0.1900
           &&
           19.566 & 0.7632 & 0.0116 & 0.3630
           \\
           &
           500
           &
           26.464 & 0.8659 & 0.0048 & 0.1820
           &&
           19.619 & 0.7634 & 0.0115 & 0.3607
           \\
           &
           1000
           &
           26.799 & 0.8669 & \underline{0.0043} & 0.1781
           &&
           19.549 & 0.7619 & 0.0117 & \underline{0.3505}
           \\
           &
           Full
           &
           26.844 & 0.8701 & 0.0044 & 0.1774
           &&
           19.617 & 0.7624 & 0.0115 & 0.3692
           \\
           \midrule
           \multirow{4}{*}{SSWL-IDN}
           &
           250
           &
           26.581 & 0.8649 & 0.0046 & 0.1793
           &&
           19.660 & 0.7645 & 0.0109 & 0.3556
           \\
           &
           500
           &
           26.778 & 0.8723 & 0.0045 & 0.1783
           &&
           19.854 & 0.7680 & 0.0107 & 0.3530
           \\
           &
           1000
           &
           \underline{27.018} & \underline{0.8744} & \underline{0.0043} & \underline{0.1732}
           &&
           \underline{20.016} & \underline{0.7706} & \underline{0.0105} & \underline{0.3505}
           \\
           &
           Full
           &
           \textbf{27.800} & \textbf{0.8815} & \textbf{0.0042} & \textbf{0.1701}
           &&
           \textbf{20.178} & \textbf{0.7739} & \textbf{0.0104} & \textbf{0.3458}
           \\
           \bottomrule
    \end{tabular}}
\end{table}

The improvements from our hybrid loss are shown in our final model, SSWL-IDN, in Table \ref{table:ssl-scores}. When compared to RVAE + SSWL, which is trained on MSE, we see imporved performance of up to 1 to 2 scores higher for both PSNR and SSIM. A full detailed ablation for our hybrid loss is available in the supplemental. 

Figs. \ref{fig:comparison-ssl-rois} and \ref{fig:comparison-preds} demonstrate precise removal of noise from whole scans as well as specific regions of interest (ROIs), proving the effectiveness of our model over baseline architectures and other self-supervised tasks. As shown, both the RVAE and SSWL are able to quantitatively and qualitatively outperform their respective counterparts.

\section{Conclusions}

We present SSWL-IDN, a self-supervised denoising model with a novel, task-relevant, and efficient surrogate task of window-level prediction. We also propose a Residual-VAE specialized for denoising, as well as a hybrid loss leveraging benefits of both perceptual and pixel-level learning. We confirm each component of our method outperforms baselines on difficult 5\% dose denoising for both in- and cross-domain evaluations, and when combined, the model significantly outperforms state-of-the-art methods. Improved denoising with limited reference data is of clinical significance to reduce harms to patients. Our future work will focus on developing cascaded and joint surrogate and downstream learning as well as 3D architectures to utilize information in the z-dimension.

\bibliographystyle{splncsnat}
\bibliography{references}

\end{document}